\documentclass[12pt]{iopart} 

\usepackage{mathptmx}
\usepackage{graphicx}

\providecommand{\mbf}{\mathbf}
\renewcommand*{\vec}[1]{\mbf{#1}}

\begin{document}

\title[]{%
$N$-tupling the capacity of each polarization state in radio links
by using electromagnetic vorticity%
}

\author{Fabrizio Tamburini$^{1,2}$\footnote{Corresponding author: tamburini@twistoff.it, fabrizio.tamburini@unipd.it}, Bo\,Thid\'e$^3$, Elettra Mari$^2$, Giuseppe Parisi$^1$, Fabio Spinello$^4$, Matteo Oldoni$^5$, Roberto A. Ravanelli$^5$, Piero Coassini$^5$, Carlo G. Someda$^1$ and Filippo Romanato$^2$} 

\address{$^1$Twistoff s.r.l., via della Croce Rossa 112, I-35129 Padova, Italy} 
\address{$^2$Department of Physics and Astronomy, University of Padova, via Marzolo 8, I-35100 Padova, Italy}
\address{$^3$Swedish Institute of Space Physics, Physics in Space, {\AA}ngstr\"{o}m Laboratory, P.\,O.~Box~537,  SE-751\,21, Sweden}
\address{$^4$Department of Information Engineering, University of Padova, via Gradenigo 5B I-35131 Padova, Italy}
\address{$^5$SIAE Microelettronica, 21, via Michelangelo Buonarroti, I-20093 Cologno Monzese, Milan, Italy}
 
\begin{abstract}
The congestion of the radio frequency bands imposes serious limitations
on the capacity and capability of modern wireless information
infrastructures. One approach to enable frequency re-use is to exploit
other physical conserved quantities of the electromagnetic fields,
such as the angular momentum in addition to linear
momentum, exploited in present-day telecommunications. Whereas in the
optical regime the increase of channel capacity by using orbital angular momentum (OAM) states was
demonstrated recently, the receiving antennas in commercial radio links
have a much smaller extent than the transmitted beam, making the signal
reception and characterization of the OAM state demanding. Moreover,
radio data transmission with $N$ channels per polarization state at the
same frequency for radio links when $N>2$ is known to be notoriously
difficult to realize even with multiport techniques, long antenna
baselines and digital post-processing. Here we report results from an
outdoor experiment where the physical properties of OAM states were used
to transfer information, using far-field multiplexing/demultiplexing
of $N=3$ coexisting, collinear, vertically polarized and mutually
independent OAM radio beams, opening new perspectives in wireless
telecommunications.
\end {abstract}

\vspace{2pc}

\maketitle

\section{Introduction}

In 1895, Guglielmo Marconi invented the wireless telegraph and from
that the communication world spread in all directions \cite{6}. In
order to allow different stations to communicate simultaneously without
interfering with each other, Marconi later proposed the rasterization
of the available radio frequency spectrum of into different sub-bands. All long-distance wireless systems presently in use exploit
the linear momentum degree of freedom of the electromagnetic (EM)
radiation, and make use of various forms of phase, frequency and/or
amplitude modulation of it. However, the ever-growing use of wireless
communication with radio unavoidably leads to the saturation of
all available frequency bands, even with the adoption of clever
engineering techniques to increase the information-carrying capacity
within a given bandwidth. The focus on the saturation problem is
by no means new, and an intense research effort is underway along
well-known avenues (\emph{e.g.}, multiport technique protocols such
as multiple-input-multiple-output, MIMO). The experimental results that
we report in this paper were obtained following a different approach
where the physical properties of electromagnetic vorticity \cite{1} were
utilized. To the best of our knowledge, the results presented here are
unprecedented at radio frequencies.

In addition to energy and linear momentum, the EM field can carry also
angular momentum. The total angular momentum, $\vec{J}=\vec{S}+\vec{L}$,
is the sum of two contributions. The quantity $\vec{S}$ represents the
spin angular momentum, which is related to photon helicity and thus
to wave polarization. As is well known, $\vec{S}$ is already being
exploited in radio communications and can yield, at most, a two-fold
increase in capacity. The other quantity, $\vec{L}$, describing the
electromagnetic vorticity of the beam, is the orbital angular momentum
(OAM) associated with the structured orbital helicoidal phase profile
that a suitable beam can exhibit, can yield an $N$-fold increase in
capacity $(N=1,2,3,\ldotsÉ)$ for each polarization state $\vec{S}$.
The OAM degree of freedom \cite{7,8,9,10} can be described in terms of
classical theory based on Maxwell's equations \cite{11,12}. Paraxial
beams carrying OAM can be readily described by Laguerre-Gaussian
(LG) modes that are identified by two integers $\ell$ and $p$. The
parameter $\ell$ describes the number of twists of the helical wavefront
in a wavelength, whereas $p$ gives the number of radial nodes of the
mode, characterized by a phase profile azimuthally spanning $2 \pi
\ell$ radians in the plane orthogonal to the beam propagation. When
OAM beams propagate, they have a characteristic intensity (linear momentum density /Poynting vector) profile with a `doughnut' shape with a null on the
propagation axis. This fundamental physical property of EM waves has
already found practical applications in several fields: nanotechnology
\cite{13,14}, optical \cite{2,3} and radio \cite{1,12} communications
and in astronomy \cite{15,16,17,18} to improve the resolving power of
diffraction-limited optical instruments \cite{19}, to image extrasolar
planets \cite{20,21,22} and to detect Kerr black holes \cite{23}.

In our experiment we achieved a stable link with three beams over a
distance of $100$ meters ($5700$ wavelengths $\lambda$ at $17.1$--$17.3$
GHz) by using only simple analog post-processing to enhance the
signal-to-noise ratio (SNR). These three independent beams were
modulated by means of $4$-Quadrature Amplitude Modulation (4-QAM)
and carried each about $11$ Mbit s$^{-1}$. Our results clearly prove
that $N>2$ OAM radio beams per polarization state can be used
for simultaneous data transmission for a given link. We also set up
a bidirectional data exchange between two OAM radio stations, each
carrying a $436$ Mbit s$^{-1}$ $1024$-QAM signal, and two simultaneous
OAM transmissions with digital modulation up to $2 \times 171$ Mbit
s$^{-1}$, opening new perspectives in wireless telecommunications.

\subsection{Schematics of the experiment}

With present technologies, the detection and characterization of EM
waves carrying OAM can, in experiments where a given OAM state is being
used, be made with simple interferometric methods, such as those used in
radio astronomy \cite{1,24}. For telecommunication purposes, however,
the practical needs of construction and installation require that both
antennas, in case of bidirectional links, have similar characteristics.
To this end, in order to transmit and receive electromagnetic waves
carrying OAM, we used pairs of identical twisted parabolic antennas
\cite{1}, each couple generating and receiving either $\ell=+1$ or
$\ell=-1$ LG-OAM modes with $p=0$. A twisted parabolic antenna is a
reflection phase mask that converts a spherical phase front into a
helical one with a specific OAM value. The reception of OAM beams is
also made through an identical twisted parabolic reflector placed in
front of the transmitter. Because of parity change (recall that angular
momentum is a pseudovector), this latter antenna acts as an inverse
phase mask and transforms the received OAM beam into a planar beam that
is injected into the feeder/coupler and propagates to the receiver via
the fundamental mode of the waveguide. If the impinging beam has instead
a different OAM $\ell$ value than the receiving parabola, their mutual
orthogonality suppresses the former, preventing it from reaching the
receiver.

Initial tests of the newly designed and constructed twisted antennas
were made at distances up to $7$ meters, in the (linear momentum)
near field zone. They showed an excellent modal insulation even when
two opposite antennas were coaxially placed face-to-face at a short
distance (a few centimetres). On the other hand, we observed good
transmission between two equal twisted antennas facing each other (see
Table \ref{tab1}).

\begin{table}
\center
\begin{tabular}{|c|c|c|c|}
\hline
 {Tx}&{Rx / status }&{Parameter}&{Measurement} (dB) \\
\hline
{$\ell=0$}&OP&RL&13.7 \\
\hline
{$\ell=0$}&SH&RL&2.3\\
\hline
{$\ell=0$}&$\ell=0$&IL&3.3\\
\hline
{$\ell=+1$}&OP&RL&14.2\\
\hline
{$\ell=+1$}&SH&RL&16.2\\
\hline
{$\ell=+1$}& $\ell=+1$&IL&3.2\\
\hline
{$\ell=+1$}& $\ell=0$&IL&18.6\\
\hline
{$\ell=+1$}& $\ell=-1$&IL&26.2\\
\hline
\end{tabular}
\caption{%
 End-to-end measurements with a radome (Tx is the transmitted beam, IL
 the insertion loss, RL the return loss, OP the open field, SH with a
 metal plane covering the whole aperture.
}
\label{tab1}
\end{table} 

\begin{figure}
\centering
 \includegraphics[width=1 \columnwidth]{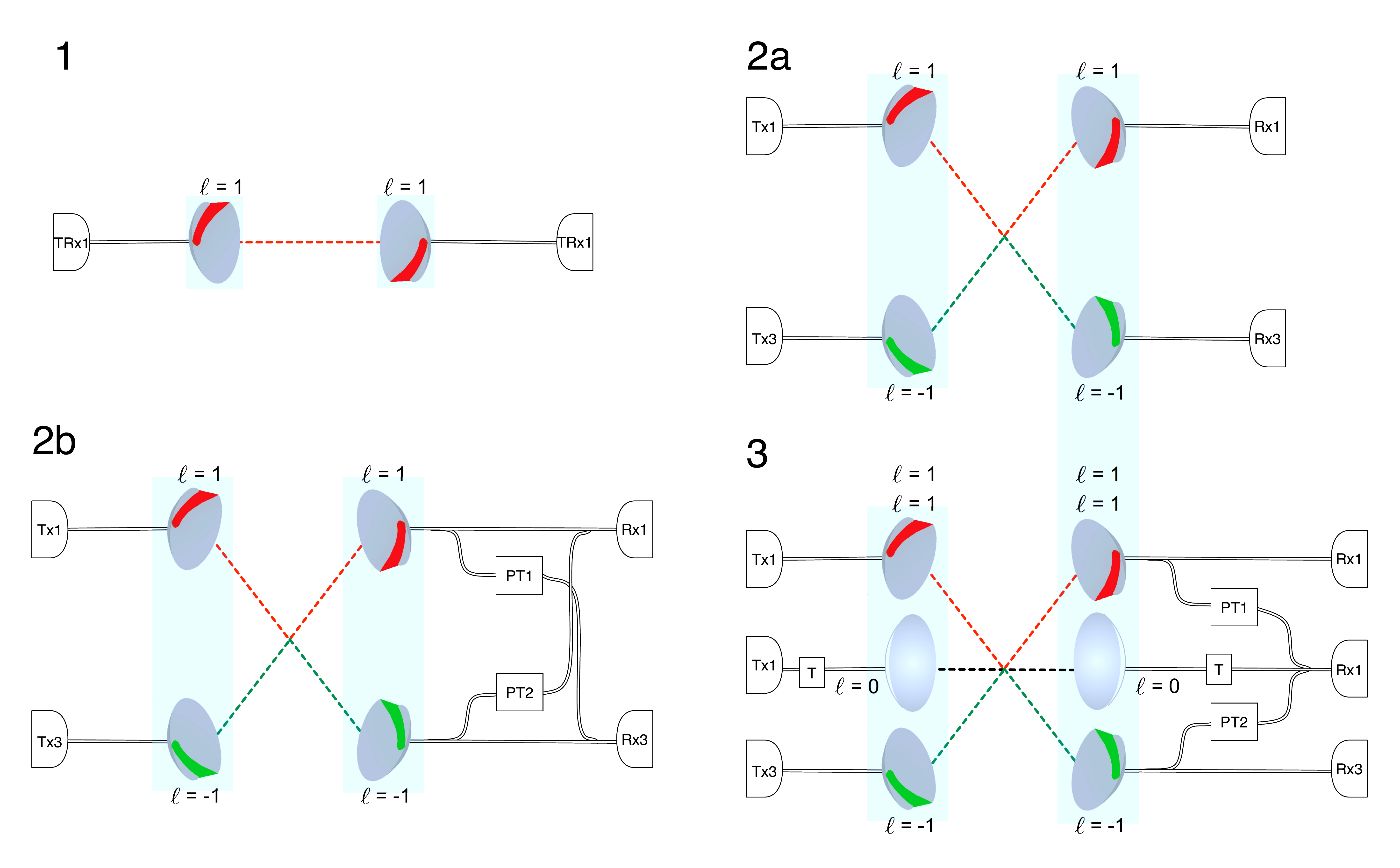}
\caption{%
 Schematics of the radio experiments. 1: single link; 2: double link
 without (a) and with (b) additional mode cancelling; 3: triple link.
 The letter T denotes 10 dB attenuators, whereas PT represents phase
 shifter and rotary attenuator blocks. Tx is the transmission unit, Rx
 the receiver. Dashed lines represent the directions where the antennas
 are pointed (angles and distances are not in scale). The bifurcation of
 each signal line represents a -3dB directional coupler.
}
\label{fig1}
\end{figure}

In order to investigate the data transport capabilities of OAM radio
beams, we designed and ran three experiments in the $17.1$--$17.3$~GHz
unlicensed band at a distance of $100$ metres with the antennas mounted
on the rooftops of two buildings of the SIAE Microelettronica industrial
compound near Milan, Italy: (1) a single bidirectional link without
additional analog interference suppression, (2) a double unidirectional link
without and with additional analog interference suppression, and (3) a triple
unidirectional link with additional analog interference suppression.
The radio equipment used in these experiments was a commercially
available outdoor transmitting/receiving unit manufactured by SIAE
Microelettronica. These units are completely configurable in their radio
properties of interest (transmitted power, modulation scheme, carrier
frequency and modulated bandwidth). A block diagram of the various
experiments is shown in Figure~\ref{fig1}, with all the antennas mounted
for operation in vertical polarization.

\begin{figure}
\centering
 \includegraphics[width=1 \columnwidth]{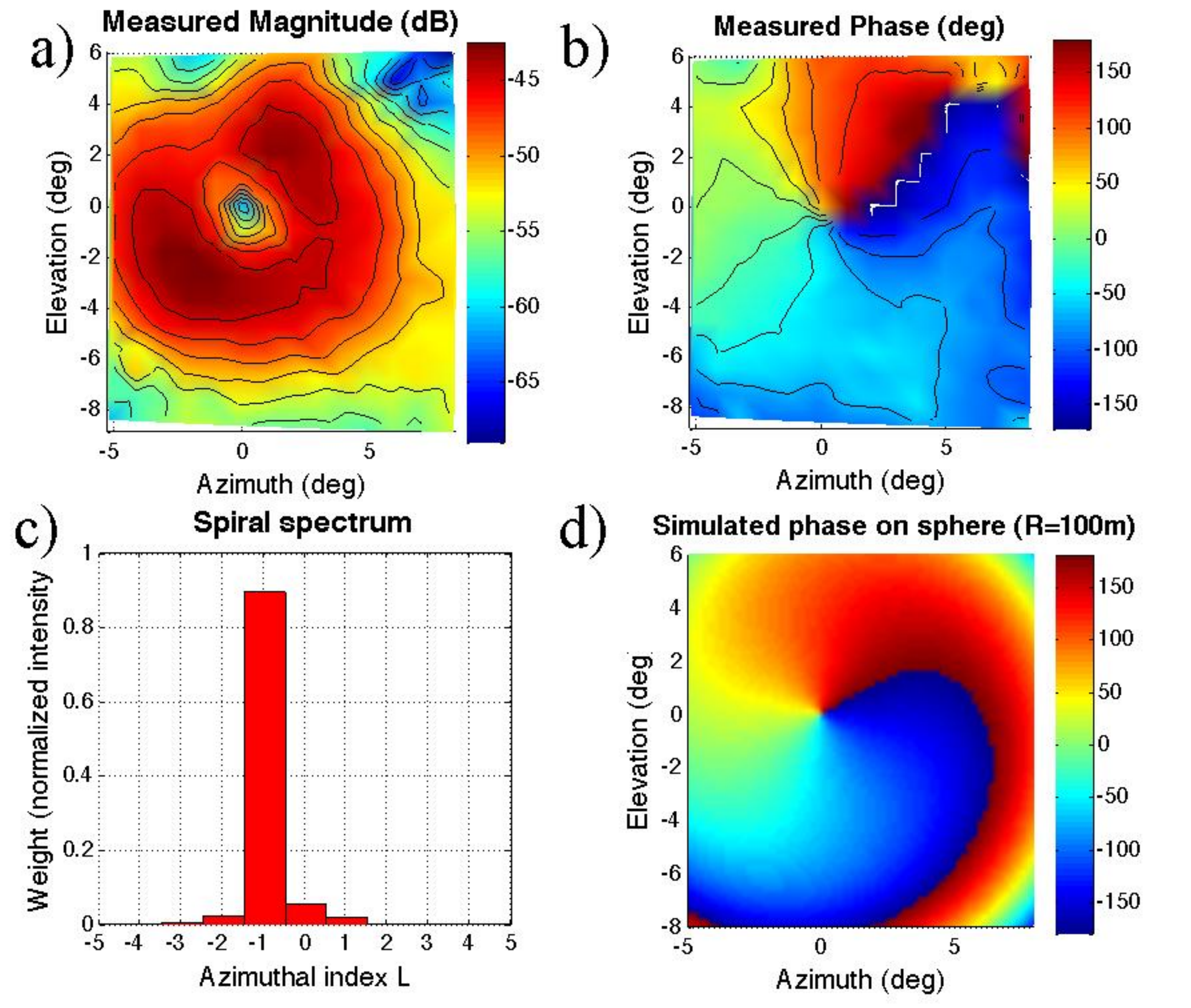}
\caption{%
 Intensity (magnitude of the linear momentum density) a) and phase
 map b) of the $\ell=-1$ radio beam vortex measured in free space at
 a distance of 100 m. The corresponding OAM spectrum, peaking at the
 $\ell=-1$ value, as expected, is shown in the inset c). The dip at
 the center of the intensity distribution corresponds to the phase
 singularity and the notch depth is 30 dB below the peak. Instead,
 d) shows the simulation of the phase distribution obtained through
 numerical simulations that confirm the good quality of the experimental
 data.
}
\label{fig2}
\end{figure}

The measured phase and intensity profiles of the OAM beams are presented
in Figure~\ref{fig2}; the received-power matrix of the setup is reported
in Table \ref{tab2}.

\begin{table}
\center
\begin{tabular}{|c|c|c|c|c|}
\hline
 &{Tx1 $\ell=+1$}&{Tx2 $\ell=0$}&{Tx3 $\ell=-1$}&{SNR} (dB) \\
\hline
{Pow} (dBm)&22&8&22& \\
\hline
{Rx1 $\ell=+1$}&-55.500&-65.625&-72.625&9.5\\
\hline
{Rx2 $\ell=0$}&-73.000&-55.375&-65.375&9.3\\
\hline
{Rx3 $\ell=-1$}&-70.125&-52.875&-40.875&11.3\\
\hline
\end{tabular}
\caption{Matrix of received powers and estimated signal-to-noise ratio (SNR) for the triple link setup. Pow indicates the output power from each radio transmitter.}
\label{tab2}
\end{table} 

The $3$ dB ring of these OAM beams were about $6$ and $9$ metres in diameter, respectively, implying that the dark region hosting the singularity was much larger than the receiving antenna cross section. Because of the doughnut geometry, one can obtain good signal and phase information by placing the receivers in an off-centre region of higher intensity (linear momentum density/Poynting vector), although this decreases the orthogonality among the received OAM modes. Additional modal isolation can be obtained by exploiting the topology of OAM beams.
What is crucial is that this situation is different from the experiments in the optical domain where most of the beam is collected; in radio links, only a small fraction of the beam can normally be gathered by the receiving antenna.

The single link experiment (Figure~\ref{fig1}, inset 1) proved that
the beam generated by the twisted $\ell=+1$ antenna can support a
complex modulation scheme ($1024$-QAM) over a wide bandwidth ($56$ MHz),
resulting in an error-free throughput of $436$~Mbit/s per direction with
a transmitted power of $17$~dBm, such as for a standard radio beam.

The double link experiment was conceived to test unidirectional
data transfer via two $20$ dBm OAM beams generated by the twisted
parabolas $\ell=\pm 1$ on the same carrier frequency ($17.128$ GHz) and
polarization; see Figure~\ref{fig1}, insets 2a and 2b. To this end we
exploited the peculiar linear momentum radiation pattern of the antennas
so that the interfering signals reaching the $\ell=\pm1$ receivers were
minimum, due to the doughnut-shaped linear momentum radiatiation pattern
and to the mutual orthogonality of the coaxial OAM beams. The error-free
throughput achieved was $2\times171$~Mbit/s, with $16$-QAM modulation
over a $56$ MHz bandwidth. We further improved the isolation between
channels by introducing a pair of attenuator-phase shifter blocks,
implementing a basic static cancelling scheme. This however required
a reduction of the bandwidth to $7$ MHz only due to the narrow-band
response of the phase shifters. The modulation was however increased
to 256-QAM, with an errorless throughput of $2\times42$~Mbit/s with
transmitted power of $18$~dBm.

\begin{figure}
\centering
  \includegraphics[width=1 \columnwidth]{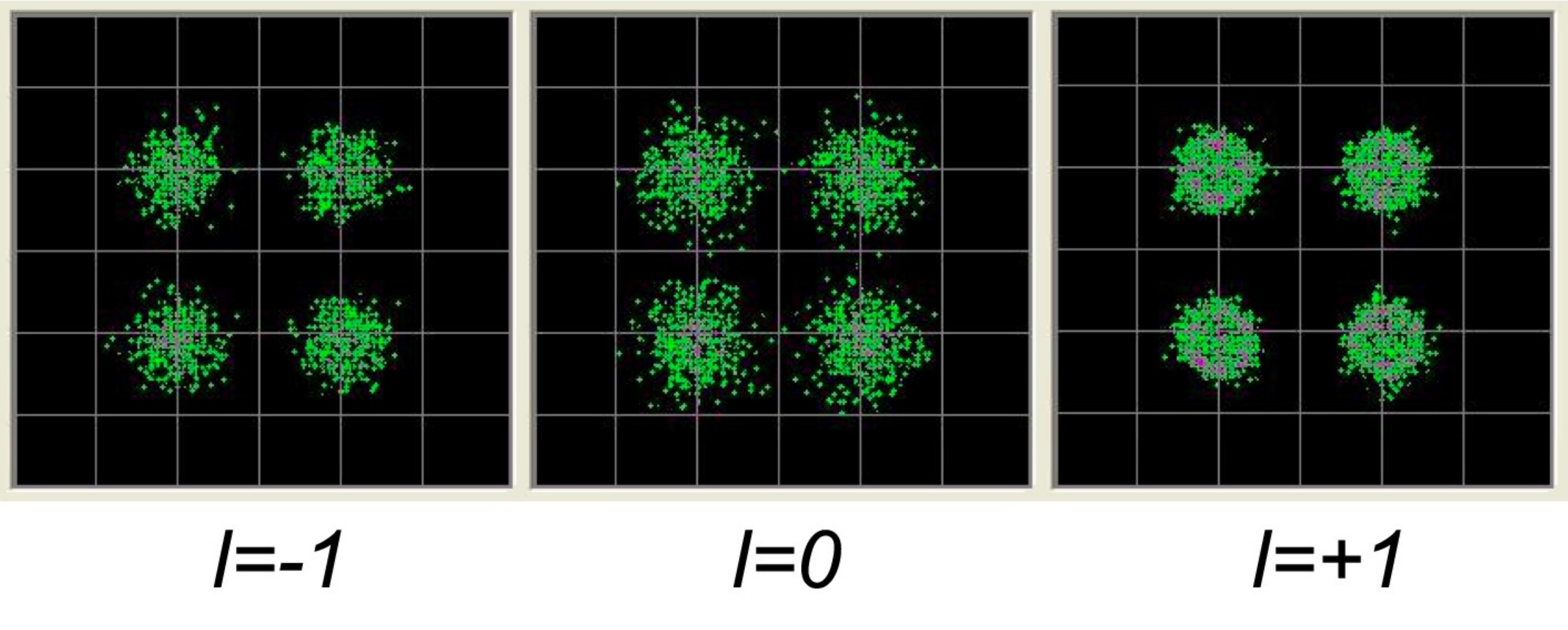}
\caption{%
 Received digital $4$-QAM constellations of three independent digital
 channels sent through as many different and superimposed OAM beams on
 the same carrier frequency, bandwidth and polarization. By using the
 topological properties of OAM beams, each receiver detects its specific
 OAM state/channel. From left to right: channels RX1 ($\ell=-1$), RX2
 ($\ell=0$) and RX3 ($\ell=+1$).
}
\label{fig3}
\end{figure}

The triple link at $17.128$ GHz was set up by adding an $\ell=0$        
(standard antenna) channel to the double link experiment. To minimize   
the interference of the transmitted signals on the relative opposite    
receivers, we exploited the aiming of the twisted parabolas as shown    
in Figure~\ref{fig1}, inset 3, so to use their inherent `natural'  
mode insulation of about $23$ dB. he pair of static mode-cancellers     
removed the interference of the transmitted signals in the $\ell=0$     
channel, due to the loss of orthogonality experienced by the standard   
antenna when receiving a portion of the radiated OAM beam far off its   
axis. With such a scheme we achieved an error-free throughput of $3     
\times 11$ Mbit/s with $4$-QAM over a $7$ MHz bandwidth, with standard  
forward error correction based on low-density parity-check codes (LDPC) 
algorithm \cite{25}. A screenshot of the received constellations        
is shown in Figure~\ref{fig3}, whereas the received powers and the        
signal-to-noise ratio (SNR) values for the triple link setup are        
reported in Table \ref{tab1}.                                           

To minimize the mutual interference on the $\ell=0$ receiver we then
aimed the beam axes of the twisted receiving parabolas toward the
untwisted transmitting antenna and increased the capacity of the
$\ell=0$ beam to $16$~QAM. We also changed the modulation bandwidths to
$56$~MHz for the twisted channels and to~$7$ MHz for the untwisted one,
respectively.

\section{Twisted parabolic antennas, design and realization}

To optimize the antenna design for both transmission and reception, we
performed a numerical study with MOM (method of moments) simulations
to couple, in the best way, the geometry of the modified parabolic
phase mask with the hyperbolic feeder from the original standard
Cassegrain antenna. The original parabolic reflector was a standard
$36$~cm parabola with approximately $30$--$35$~dB gain at $17.2$~GHz. A
conventional parabolic reflector described in a cylindrical coordinate
system $(r,z,\varphi)$ ideally produces plane wavefronts at very long
distances from a spherical source at the focus thanks to its profile $z$
as a function of distance, $r$, from its axis. For a focal length $F$
the profile is

\begin{equation}
z(r)=\frac{r^2}{4F}
\label{eq1}
\end{equation}

To imprint vorticity onto the beam emitted by the Cassegrain feeder, one
can deform the shape of the standard parabolic reflector and azimuthally
gradually elevate its surface along the symmetry axis as a function
of the azimuthal angle $\varphi$ so as to obtain a final step of half
the wavelength after having encircled the symmetry axis an angle of
$\varphi=2\pi$. The modified parabolic reflector designed to generate a
vortex with order $\ell=m$ is defined by the formula
\begin{equation}
\label{eq2}
 z(r,\varphi)=\frac{m(\varphi-\pi)\lambda}{4\pi}
 + \frac{\pi r^2}{4\pi F - m(\varphi - \pi)\lambda}.
\end{equation}
Equation \ref{eq2} ensures that the focus $F$ of the twisted reflector
is independent of $\varphi$ and that the focus of the twisted parabolic
antenna coincides with the focus $F$ of the original parabolic
reflector. Our twisted antennas have a gap of $\lambda/2$ at $17.2$~GHz
($\lambda=1.74$ cm).

\begin{figure}
\centering
 \includegraphics[width=1 \columnwidth]{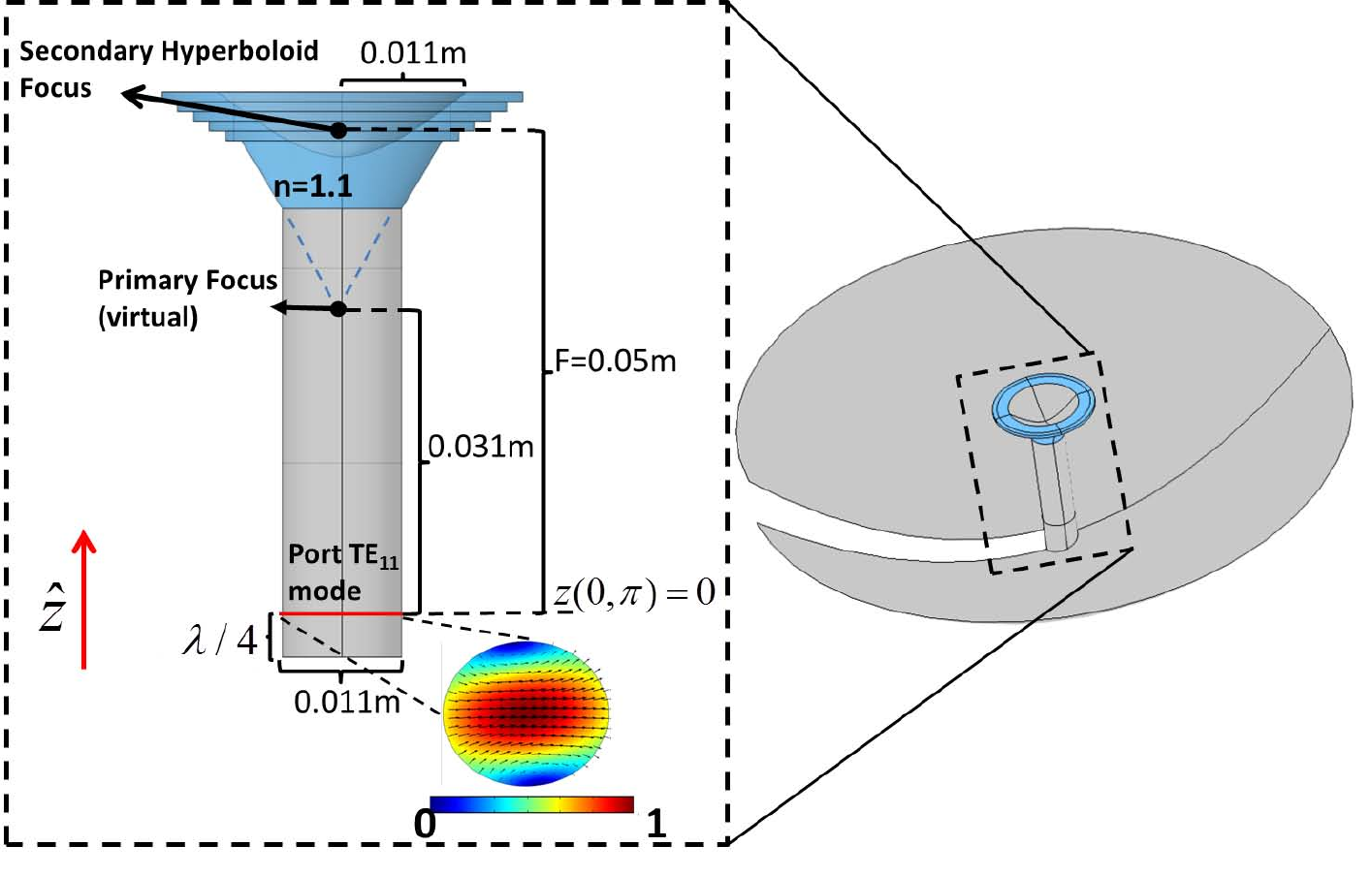}
\caption{%
 Layout of the Cassegrain antenna configuration for the twisted
 ($\ell=+1$) parabolic reflector and the (transversal electric) TE11 mode
 used as port excitation in the simulation, intensity (color plot) and
 direction (arrows).
}
\label{fig4}
\end{figure}

Figure~\ref{fig4} shows the Cassegrain configuration of the twisted
parabolic antenna. The feeder is mounted in the same position as for
a conventional parabolic Cassegrain antenna. Both the left-handed
($\ell=+1$) and the right-handed ($\ell=-1$) OAM mode antennas as well
as the untwisted one ($\ell=0$) have the same focus.

Numerical simulations show that the feeder emits a wave resembling the
(transversal electric) TE11 waveguide mode. The MOM simulations of
the complete structure (feeder plus twisted parabolic reflector) were
carried out by setting the TE11 waveguide mode as port excitation.

\begin{figure}
\centering
 \includegraphics[width=1 \columnwidth]{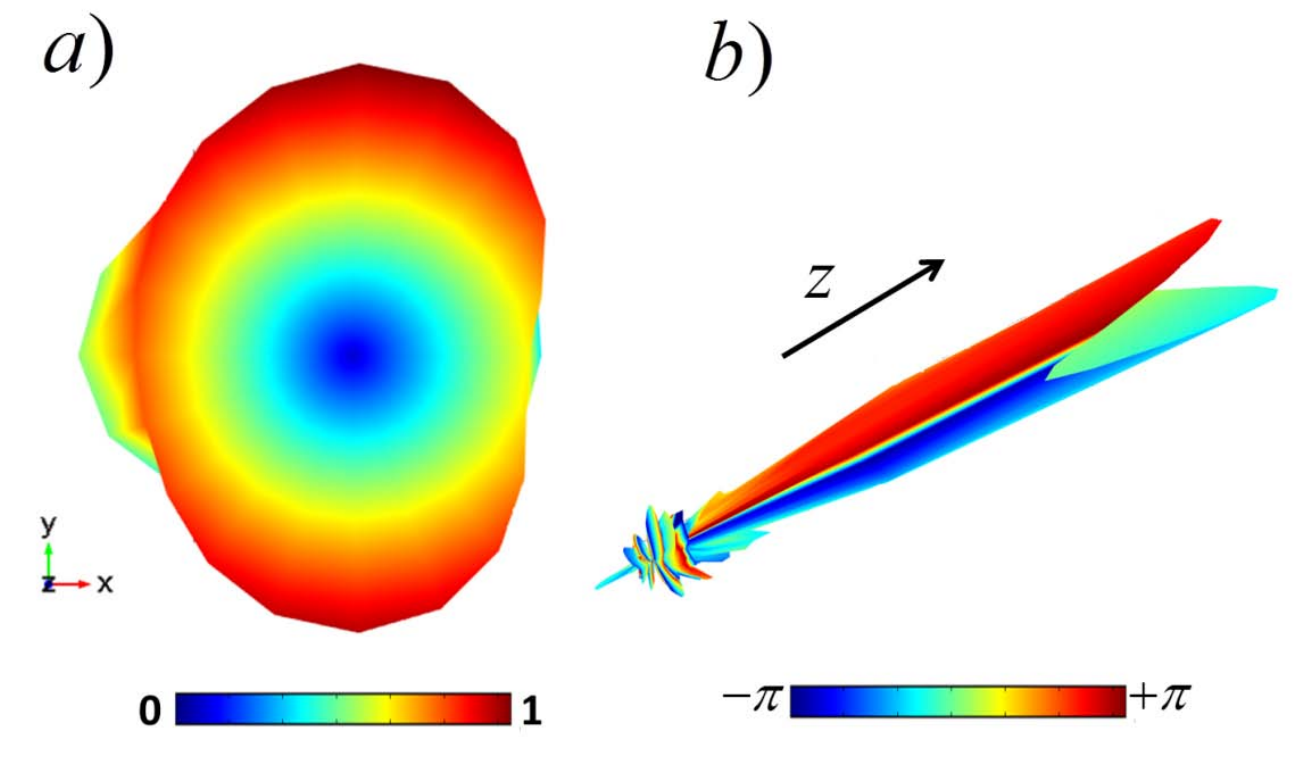}
\caption{%
 a) The magnitude of the electric field in the far field zone calculated
 in the plane perpendicular to the propagation direction. The scale is
 in arbitrary units. b) Three dimensional plot of the radiation lobe
 emitted by the twisted parabolic reflector. Here, the color palette
 represents the phase of the x-component of electric field, calculated as
 $\arg(E_x)$.
}
\label{fig5}
\end{figure}

Figure~\ref{fig5} shows the radiated electric field as obtained in the
MOM simulations in the far zone of a twisted antenna. The electric
field magnitude and phase exhibit the expected doughnut shape together
with the phase variation and the singularity located at the centre. The
phase of the polarized electric field is obtained by estimating the
argument of the complex function describing the electric field on its
polarization axis, $\arg(E_x)$, and the numerical simulations confirm
that the vortex emitted by the antenna has the typical $0$--$2\pi$
phase variation of the $\ell=1$ mode. The peculiar `fish-mouth' shape
exhibited by the radiated vortex, as shown in the right panel b) of
Figure~\ref{fig5}, is caused by the two local minima present in the TE11
mode coming from the circular feeder propagating in the far field
region.

\subsection{Realization and test of the modified parabolic antennas}

The twisted parabolic antennas were manufactured by using the
stereolithography printing technique (U.S. Patent 4,575,330) directly
from 3D CAD plots used in our MOM simulations. The polymer-derived
ceramic support was 3 mm thick with a surface roughness of $0.1$~mm
thick and the inner part of the antenna was coated with a 0.1
mm-thick copper layer deposited with a galvanic method with a $0.25\%$
surface-error tolerance.

\begin{figure}
\centering
 \includegraphics[width=1 \columnwidth]{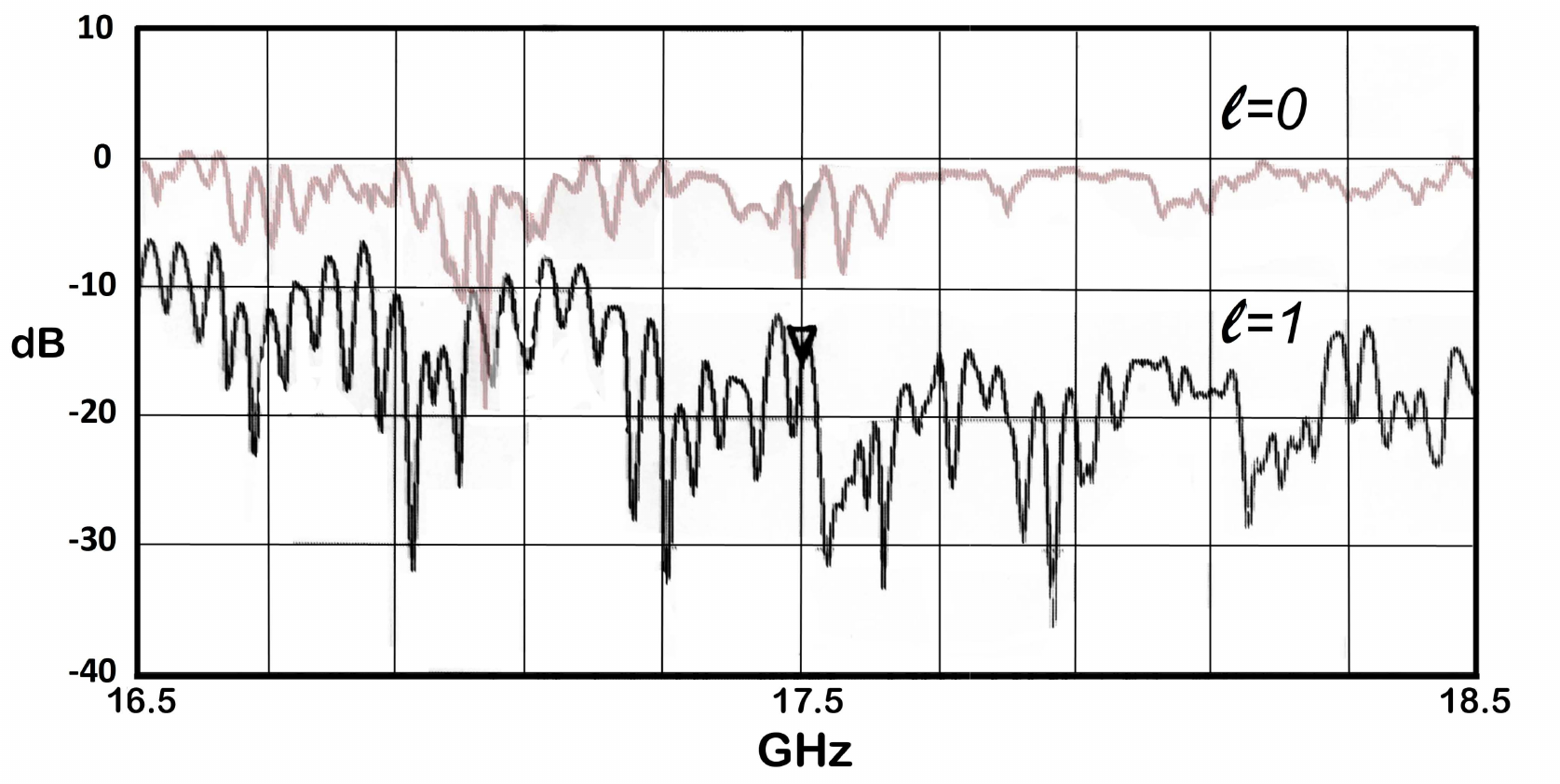}
\caption{%
 Return loss (RL) of a standard parabolic antenna ($\ell=0$) with
 respect to a twisted one ($\ell=1$) when both were placed in front of
 a metallic surface. The standard antenna is sensitive to its reflected
 wave and therefore exhibits a poor RL; by contrast, the OAM antenna,
 thanks to the parity change, is not sensitive to its reflected wave.
}
\label{fig6}
\end{figure}

The electrical parameters of these antennas were measured by evaluating
the short-circuit return loss (RL) in open field (OP) and with a metal
plane covering the whole aperture (SH). We measured the ratio between
the reflected wave coming from the antenna port and the signal injected
into it when the antenna was located in front of a metallic surface.
The OAM and transient modes, present in the near field, because of the
change of parity due to the reflection on the metallic surface, actually
make the twisted antenna exhibit a rather different behavior compared to
a standard parabolic antenna, as depicted in Figure~\ref{fig6}.

We also evaluated the transmission characteristics (insertion loss,
IL) for various combinations of transmitting and receiving antennas
when facing each other at a short distance ($< 1$ cm between the planes
of the apertures) and with and without a protective radome cover.
Tables \ref{tab1} and \ref{tab3} report the measured values at $17.2$
GHz. To analyze the results, we used the standard parabolic reflector
($\ell=0$) as reference (first three rows in Table \ref{tab1}). The RL
is acceptable ($13$ dB) when the antenna is free to radiate, whereas it
drops dramatically to $2.3$ dB when the aperture is closed by a brass
plate. This is a clear indication of an almost complete rejection of
the radiated power, being reflected back into the feeder and on to the
generator. We also evaluated the transmission between homologue antennas
at a distance of 2 m (see Table \ref{tab4}).

\begin{table}[htbp]
\center
\begin{tabular}{|c|c|c|c|}
\hline
 {Tx}&{Rx / status }&{Parameter}&{Measurement} (dB) \\
\hline
{$\ell=+1$}&OP&RL&13.4 \\
\hline
{$\ell=+1$}&SH&RL&15.5\\
\hline
{$\ell=+1$}&$\ell=+1$&IL&3.1\\
\hline
{$\ell=-1$}&$\ell=+1$&IL&16.8\\
\hline
\end{tabular}
\caption{End-to-end measurements without radome (IL = insertion loss, RL = return loss, OP = open field, SH = with a metal plane covering the whole aperture.}
\label{tab3}
\end{table} 

\begin{table}[htbp]
\center
\begin{tabular}{|c|c|c|c|}
\hline
 {Tx}&{Rx / status }&{Parameter}&{Measurement} (dB) \\
\hline
{$\ell=0$}&$\ell=0$&IL&7.74 \\
\hline
{$\ell=+1$}&$\ell=+1$&IL&12.24\\
\hline
\end{tabular}
\caption{Transmission at 2 m, with radome (IL = insertion loss).}
\label{tab4}
\end{table} 

The same measurements for a twisted parabola $\ell=+1$ show a good RL
($14.2$ dB) for unimpaired radiation, thus suggesting that the antenna
is indeed radiating power. Interestingly, a similar value ($16.2$ dB) is
also exhibited as RL when the aperture is closed onto the brass plate.
We explain this phenomenon by the behavior of OAM beams reflected from
perfect electric conductors: the wave changes its topological charge
due to a reversed propagation direction. Therefore, it impinges as an
$\ell=-1$ beam onto the $\ell=+1$ antenna that produced it and thus,
due to the opposite curvature of the reflector, does not reach the
generator. As a consequence, the wave remains ÒtrappedÓ within the
cavity and is dissipated by the conduction losses of the metal surfaces.

This behaviour is virtually unaffected by the presence of the
radome cover, as can be seen by comparison with Table \ref{tab1}.
Concerning the transmission characteristics, we observe that a pair of
$\ell=0$ antennas exhibit an IL value of about 3 dB, which, for this
experiment, is thus considered, to be a good power transfer. A pair
of $\ell=+1$ antennas show similar behavior, suggesting that the same
power flow is sustained between them. This may however raise doubts
concerning the validity of such measurements, as we are located well
into the near-field region. If the antennas are generating a residual
$\ell=0$ beam, this might transfer power between them. But a further IL
measurement between $\ell=+1$ and $\ell=0$ shows that the transmission
is severely impaired, and similarly between $\ell=+1$ and $\ell= -1$. We
deem this to be compatible with the hypothesis of the receiving antenna
being sensitive to a perpendicular beam, as happens with two antennas
with orthogonal polarizations. Transmission among homologue antennas at
a distance of $2$ m ($\sim 110$ wavelengths) shows that the magnitude
of the received signal in the $\ell=+1$ pair is about 4.5 dB smaller
than the corresponding $\ell=0$ pair. This can be ascribed to the larger
spread of the $\ell=+1$ beam, which is thus captured to a lesser degree
by the finite area of the receiving antenna than the $\ell=0$ beam.

\section{OAM radio links---experimental results}

In this Section we present a more detailed account of the experimental
results of transmission/reception for each link pictured in
Figure~\ref{fig1}.

\paragraph{1. Single OAM link:}
Two twisted $\ell=+1$ antennas, manufactured and characterized as
described in the previous section, were deployed at the two sites
(namely A and B), respectively, and connected to a pair of commercially
available radio Full Outdoor Units provided by SIAE Microelettronica.
They function as full duplex transceivers in the chosen frequency band
and implement standard forward error correction based on the LDPC algorithm.
The frequency for the A $\rightarrow$ B link is $17.128$ GHz whereas for
the B $\rightarrow$ A direction it is $17.272$ GHz. The $\ell=+1$ at
site A (B) was aimed towards its counterpart at site B (A) which caused,
as a side effect, the received power to lie in a relative minimum due to
the doughnut-shaped radiation pattern. We set up the transceiver units
to operate with a modulation bandwidth of $56$ MHz and increased the
constellation levels up to $512$-QAM, which carries $9$ bits/symbol. By
tilting the antennas slightly upward ($<1$ degree), we were moreover
able to increase the received power and safely switch to a $1024$-QAM,
which carries $10$ bits/symbol. This tilt is not strictly necessary to
operate the link, but may be used as a way to overcome the minimum of
received signal with direct aiming. This is consistent with the measured
radiation pattern of these twisted parabolas, whose intensity ÒdoughnutÓ
at the selected distance exhibits a radiation maximum at an angle
between $2$ and $3$ degrees from the axis. The number of errors detected
at both sides remained zero for the whole duration of the experiment
(one day), and in fact the evaluated signal-to-noise ratios were both
well above the equipmentÕs error-free safety thresholds. The total
throughput for this duplex link was about $2 \times 436$ Mbit/s.

\paragraph{2. Double OAM link:}
A setup similar to the one described for the single link was used here,
except that in this case we mounted an $\ell=+1$ and an $\ell=-1$
antenna on the masts at both sites of the link, in order to set up a double
unidirectional link from site A toward site B on a single frequency
($17.128$~GHz) and with the same polarization (vertical for both
antennas). The antennas on the mast were spaced by about $1$ m between
their apertures. We first aimed the two transmitting antennas (at site
A), so that their axes pointed approximately towards their opposite
counterpart. In other words, $\ell=\pm 1$ at site A were aimed toward
$\ell= \mp 1$ at site B, in a sort of cross-aiming manner to minimize the
received interference power at the receivers. Then, the aiming of the
four antennas was slightly adjusted to provide the minimum received
power at $\ell=\mp 1$ due to the transmitting $\ell=\pm1$, switched
on one at a time. Once a proper aiming was obtained, we switched
both transmitters on and increased the bandwidth and modulation up
to $56$ MHz with a $16$-QAM, carrying 4 bits/symbol and achieving an
error-free throughput of $2 \times 171$ Mbit/s. In order to improve
the signal-to-noise ratio at the receivers, we then introduced an
elementary static analog cancellation scheme, based on a pair of rotary
attenuator-phase shifter blocks. These were wired at the receiver end
via 3 additional directional couplers so that the residual $\ell=\mp
1$ signal received spuriously by $\ell=\pm1$ could be suppressed. We
manually adjusted the attenuation levels and phase delays to reduce this
interference and achieved an error-free $256$-QAM modulation (carrying
$8$ bits/symbol). The total throughput was $2\times42$ Mbit/s, as
in fact we used a smaller bandwidth ($7$~MHz) to match the inherently
narrow-band response of the phase shifters.

\paragraph{3. Triple link:}
To further investigate the capabilities of multiple OAM beams on
the same frequency and with the same polarization, we installed an
additional standard ($\ell=0$) parabolic antenna at each site of the
double link in order to establish a triple link A $\rightarrow$ B
at 17.128 GHz with vertical polarization. The standard antennas had
a diameter of 36 cm, equal to the maximum diameter of the twisted
parabolas and had same focal ratio. The three antennas at each site
were vertically stacked with a distance of about 30 cm between their
apertures. The $\ell=0$ antennas were equipped with an additional
$10$ dB attenuator so that the power levels at the three receivers
due to their respective transmitters were comparable. For the same
purpose we also calibrated the transmitters to output different power
levels, as reported in Table~\ref{tab2}. The two $\ell=0$ antennas
were directly aimed toward each other. The first tests were made by
aiming the transmitting twisted antennas to those with opposite OAM
sign, obtaining a $3$ channel $4$-QAM link. To optimize the link, the
receiving $\ell=\pm1$ at site B were then aimed so that the power
received by them due to the transmitting $\ell=0$ at site A was minimum.

Next we cross-aimed the remaining antennas in a manner similar to that described for the double link (without the additional mode suppression): the $\ell=\pm 1$ antennas at site A were rotated so that their opposite counterparts ($\ell=\mp 1$ at site B) received the minimum interfering signal. Such an aiming procedure relies on two of the main properties of OAM beams:
\begin{itemize}
 \item 
 In order to have minimum interference on $\ell=\pm1$ at site B due
 to the $\ell=0$ beam, we aimed the receiving OAM antennas toward the
 source of $\ell=0$ so that they intercepted a portion of a standard
 Gaussian beam, which can also be assumed locally and at large distance
 to be a plane wave. Thanks to the intrinsic orthogonality of OAM
 beams with plane waves, the $\ell=\pm 1$ at site B will theoretically
 transfer no $\ell=0$ power to their receivers. In this way we increased
 the capacity of the $\ell=0$ beam to $16$ QAM, and widened the
 modulation bandwidths to $56$~MHz for the twisted channels and $7$~MHz
 for the untwisted one, respectively.

 \item
 In order to have minimum interference on $\ell=\pm 1$ at site B due
 to the $\ell=\mp 1$ at site A, we exploited in this case mainly the
 radiation zero, \emph{i.e.} the doughnut centre exhibited by the
 transmitting antennas. In fact, in the double link experiment, we
 reciprocally cross-aimed the antennas at both sides and were thus
 able to place the receiver antenna at the same spot but we 
 also experienced the intrinsic orthogonality of coaxial OAM beams.
 In the triple link experiment, however, the receiving antenna was
 not pointing exactly toward the opposite OAM source and therefore it
 cannot, analytically, rely on a complete modal orthogonality. The
 radiation zero, however, manifests itself as a notch with extremely
 weak fields and, the receiving antenna being small with respect to
 the size of this dip, it is sufficient to provide an acceptable
 isolation.

\end{itemize}

Once a proper aiming had been obtained, we introduced a basic static
cancelling scheme similar to the one described for the double link case,
although here wired to cancel the residual $\ell=\pm1$ interference onto
the $\ell=0$ branch. This is necessary since the $\ell=0$ antenna is
sampling only a portion of the OAM beam and thus cannot fully utilize
the modal orthogonality. Hence, at its port, the $\ell=0$ is affected
by heavy interference that must be reduced in order to achieve an
acceptable signal-to-noise ratio.

\section{Conclusions}

This experiment verifies the basic properties of radio OAM
beams and provides a practical proof of the data transport
characteristics of OAM beams. It also constitutes an example of radio
OAM multiplexing/demultiplexing. The OAM therefore can be used as a
physical layer to use for increasing the capacity of radio links and
communication systems, also when only a small fraction of the EM beam is
received, a rather common situation in real-world radio communication
scenarios. Further theoretical and experimental investigations are
called for.

\section*{Acknowledgments}
The authors acknowledge the logistic and financial support of SIAE
Microelectronics in the designing, building, and testing of the setup.
B.T. also gratefully acknowledges the financial support from the Swedish
National Space Board and the Swedish Research Council under the contract number 2012-3297.

\end{document}